\documentclass[
  aps,
  prb,
  twocolumn,
  superscriptaddress,
  floatfix,
  nofootinbib,
  amsmath,
  amssymb,
  showpacs
]{revtex4-2}

\usepackage{bm}
\usepackage{graphicx}
\usepackage{xcolor}
\usepackage{hyperref}
\hypersetup{colorlinks=true,
            linkcolor=blue!45!black,
            citecolor=blue!45!black,
            urlcolor=blue!45!black}

\newcommand{\ii}{\mathrm{i}}
\newcommand{\dd}{\mathrm{d}}
\newcommand{\Real}{\operatorname{Re}}

\newcommand{\abinitio}{\textit{ab initio}}

\newcommand{\HAIM}{\hat{H}_{\mathrm{AIM}}}
\newcommand{\Himp}{\hat{H}_{\mathrm{imp}}}
\newcommand{\Hbath}{\hat{H}_{\mathrm{bath}}}
\newcommand{\Hhyb}{\hat{H}_{\mathrm{hyb}}}
\newcommand{\HKS}{\hat{H}_{\mathrm{KS}}}

\newcommand{\dan}[2]{\hat{d}^{\phantom{\dagger}}_{#1#2}}
\newcommand{\dcr}[2]{\hat{d}^{\dagger}_{#1#2}}

\newcommand{\nop}[2]{\hat{n}^{\phantom{\dagger}}_{#1#2}}

\newcommand{\epsimp}[1]{\varepsilon^{\mathrm{imp}}_{#1}}
\newcommand{\epsbath}[1]{\varepsilon^{\mathrm{bath}}_{#1}}

\newcommand{\Norb}{N_{\mathrm{orb}}}
\newcommand{\Nbath}{N_{\mathrm{bath}}}
\newcommand{\Nq}{N_{\mathrm{q}}}

\newcommand{\sect}[1]{\mathcal{H}^{#1}}
\newcommand{\HilN}{\sect{N}}

\newcommand{\GSex}{\lvert \Psi_{0}^{N} \rangle}
\newcommand{\GSq}{\lvert \tilde{\Psi}_{0}^{N} \rangle}
\newcommand{\exNp}[1]{\lvert \Psi_{#1}^{N+1} \rangle}
\newcommand{\exNm}[1]{\lvert \Psi_{#1}^{N-1} \rangle}
\newcommand{\trial}{\lvert \psi_{\mathrm{trial}} \rangle}
\newcommand{\conf}[1]{\lvert #1 \rangle}

\newcommand{\EGS}{E_{0}^{N}}
\newcommand{\ENp}[1]{E_{#1}^{N+1}}
\newcommand{\ENm}[1]{E_{#1}^{N-1}}

\newcommand{\SN}{\mathcal{S}^{N}}
\newcommand{\SNp}{\mathcal{S}^{N+1}}
\newcommand{\SNm}{\mathcal{S}^{N-1}}
\newcommand{\SNpm}{\mathcal{S}^{N\pm1}}
\newcommand{\spn}[1]{\operatorname{span}\!\left(#1\right)}

\newcommand{\Gret}[1]{G^{R}_{#1}}
\newcommand{\Aspec}[1]{A_{#1}}

\newcommand{\Aplus}[1]{A^{(+)}_{#1}}
\newcommand{\Aminus}[1]{A^{(-)}_{#1}}

\newcommand{\broad}{\eta}
\newcommand{\chempot}{\mu}

\newcommand{\wplus}[2]{w^{(+)}_{#1,#2}}
\newcommand{\wminus}[2]{w^{(-)}_{#1,#2}}

\newcommand{\Mkeep}{M}

\begin{document}

\title{Quantum impurity solver for DFT+DMFT: real-frequency spectral function of Fe-heme on a trapped-ion quantum processor}

\author{Juan~W.~Pedersen}
\affiliation{Quantinuum K.K., Otemachi Financial City Grand Cube 3F, 1-9-2 Otemachi, Chiyoda-ku, Tokyo, Japan}

\author{Evgeny~Plekhanov}
\affiliation{Quantinuum, Terrington House, 13-15 Hills Road, Cambridge CB2 1NL, UK}

\author{Kazuhiro~Seki}
\affiliation{Quantum Computational Science Research Team, RIKEN Center for Quantum Computing (RQC), Wako, Saitama 351-0198, Japan}

\author{Kentaro~Yamamoto}
\affiliation{Quantinuum K.K., Otemachi Financial City Grand Cube 3F, 1-9-2 Otemachi, Chiyoda-ku, Tokyo, Japan}

\author{David~Mu\~{n}oz~Ramo}
\affiliation{Quantinuum, Terrington House, 13-15 Hills Road, Cambridge CB2 1NL, UK}

\author{Seiji~Yunoki}%
\affiliation{Quantum Computational Science Research Team, RIKEN Center for Quantum Computing (RQC), Wako, Saitama 351-0198, Japan}
\affiliation{Computational Quantum Matter Research Team, RIKEN Center for Emergent Matter Science (CEMS), Wako, Saitama 351-0198, Japan}
\affiliation{Computational Materials Science Research Team, RIKEN Center for Computational Science (R-CCS), Kobe, Hyogo 650-0047, Japan}
\affiliation{Computational Condensed Matter Physics Laboratory,
RIKEN Pioneering Research Institute (PRI), Wako, Saitama 351-0198, Japan}

\begin{abstract}
We apply quantum-selected configuration interaction (QSCI), closely related to sample-based quantum diagonalization (SQD), to an \abinitio{} multiorbital quantum impurity problem and compute its real-frequency spectral function and self-energy.
The target system is the heme $b$ cofactor of hemoglobin, an iron-porphyrin complex for which density functional approximations yield inconsistent spin-state orderings.
The impurity model comprises the full iron $3d$ shell and five bath sites and is encoded onto $20$ qubits of Quantinuum Reimei, a 56-qubit trapped-ion quantum computer.
Quantum sampling, followed by postselection and augmentation under time reversal, yields a subspace of $444$ determinants, representing approximately $1\%$ of the target particle-number sector.
Classical diagonalization within this subspace recovers the ground state with a fidelity of $0.982$ relative to exact diagonalization.
Using the resulting ground state and classically constructed particle-addition and particle-removal subspaces, we obtain the spectral function both on the real and imaginary frequency axis, without analytic continuation, in good quantitative agreement with exact diagonalization.
We also compute the impurity self-energy, a key quantity required for incorporating this solver into a self-consistent dynamical mean-field theory calculation.
Applying the Dyson equation to the impurity and bath orbitals together makes this self-energy causal by construction. By using DFT+DMFT, we identify heme as a charge-transfer insulator.
\end{abstract}

\maketitle

\section{Introduction}
\label{sec:intro}

Density functional theory (DFT) provides an exact framework for the ground-state properties, but its practical accuracy depends on approximations to the unknown exchange-correlation functional.
For strongly correlated systems, commonly used approximations can fail qualitatively, and a description of their electronic structure requires explicit treatment of dynamical correlations responsible for features such as Hubbard bands and satellites, as well as fluctuating local moments.
Dynamical mean-field theory (DMFT) treats local correlations nonperturbatively through a frequency-dependent self-energy, obtained by mapping the interacting lattice problem onto a self-consistently determined Anderson impurity model (AIM)~\cite{Anderson1961,Metzner1989,Georges1996}.
Combining DFT with DMFT yields an established \abinitio{} framework for describing correlated materials~\cite{Kotliar2006}.
A major computational bottleneck is the impurity solver, which must be invoked repeatedly throughout the self-consistency cycle.

Two widely used impurity solvers, continuous-time quantum Monte Carlo (CT-QMC) and exact diagonalization (ED), face distinct limitations.
CT-QMC is numerically exact in principle~\cite{Werner2006,Gull2011}, but can suffer from a severe fermionic sign problem in multiorbital systems at low temperatures, depending on the interaction terms and orbital basis.
Moreover, it yields imaginary-time or Matsubara-frequency data, from which real-frequency spectra must be reconstructed through analytic continuation, an ill-posed inverse problem that limits the reliable resolution of spectral features~\cite{Gubernatis1991}.
ED with a discretized bath~\cite{Caffarel1994,Capone2007,Liebsch2012} avoids analytic continuation by evaluating the Green's function through its Lehmann representation, both at Matsubara frequencies for the self-consistency cycle and at real frequencies for spectral analysis.
Its main limitation is the exponential growth of the many-body Hilbert space with the number of correlated orbitals and bath sites.
This computational bottleneck has motivated approaches that restrict the many-body configuration space, including truncated configuration interaction~\cite{Zgid2012} and methods based on natural orbitals~\cite{Lu2014,Lu2019}.

Quantum processors offer a route to solving impurity problems without explicitly storing the full many-body wave function: $\Nq$ qubits encode a Fock space of dimension $2^{\Nq}$.
The repeated solution of an interacting impurity model within a classical self-consistency loop makes DMFT a natural setting for hybrid quantum-classical computation.
Quantum algorithms for this purpose have been proposed and explored~\cite{Bauer2016,Kreula2016}, and two-site DMFT models have been studied on superconducting and trapped-ion processors~\cite{Keen2020,Jaderberg2020,Rungger2019,Ehrlich2025}.
Several approaches to evaluating impurity Green's functions on near-term quantum devices have also been developed~\cite{Endo2020,Kosugi2020,Baker2021}.
More recently, an impurity solver based on the Lehmann representation has been implemented within a DFT+DMFT cycle for a cuprate using up to $14$ qubits~\cite{Selisko2025}.
In parallel, quantum-centric supercomputing workflows have combined quantum processors with classical supercomputers~\cite{RobledoMoreno2025,Shirakawa2025}, while quantum subspace methods have been extended to excited states and Krylov constructions~\cite{Barison2025,Yoshioka2025}.
A relevant approach is quantum-selected configuration interaction (QSCI)~\cite{Kanno2023}, closely related to sample-based quantum diagonalization (SQD)~\cite{RobledoMoreno2025}.
In QSCI, an approximate quantum state is prepared and measured in the computational basis to select Slater determinants, and the Hamiltonian projected onto their span is diagonalized classically. The resulting ground state can then be used as the starting point for a Lehmann-representation calculation of the impurity Green's function.
This approach retains the classical evaluation of the Green's function while replacing the full configuration space of ED with a subspace selected by quantum measurements.
Device noise affects the selection of determinants, whereas the Hamiltonian matrix elements and the expansion coefficients within the selected subspace are computed classically.
Consequently, the lowest eigenvalue of the projected Hamiltonian remains a variational upper bound to the ground-state energy in the target symmetry sector, provided that the selected determinants belong to that sector.

Here we combine these elements to solve an \abinitio{} multiorbital impurity problem using quantum hardware and compute its real-frequency spectral function and impurity self-energy, the latter being a key ingredient of DMFT self-consistency.
Our target is the heme $b$ cofactor of hemoglobin, an iron-porphyrin complex with a challenging correlated electronic structure.
The difficulty is illustrated by the low-lying $S=0$, $1$, and $2$ states of ferrous porphyrin, whose energetic ordering depends strongly on the exchange-correlation functional.
A recent benchmark found that most functionals predict a triplet ground state, whereas the CASPT2 reference favors a quintet, with errors in relative spin-state energies reaching tens of kcal/mol~\cite{PorphyrinBenchmark2023}.
The challenge extends beyond DFT: CCSD(T) and CASPT2 disagree on some spin-state energy gaps, and achieving convergence in relative energies can require correlating roughly a hundred electrons in hundreds of orbitals~\cite{LiManni2018,Zhou2019}.
To retain all five iron $3d$ orbitals explicitly, we use a correlated subspace with $\Norb = 5$.
Including five bath sites gives a $20$-qubit impurity model whose target particle-number sector remains accessible to ED, enabling direct validation of the ground state, spectral function, and self-energy.

We solve this model using quantum sampling on Quantinuum Reimei, a 56-qubit trapped-ion quantum computer installed at RIKEN~\cite{QuantinuumRIKENH22026}, followed by classical subspace diagonalization.
Quantum sampling, postselection, and augmentation under time reversal yield a subspace of $444$ determinants, representing approximately $1\%$ of the target particle-number sector.
Within this subspace, we recover the ground state with a fidelity of $0.982$ and an energy error of $49$~meV relative to ED.
Whereas Ref.~\cite{Selisko2025} implemented a Lehmann-based impurity solver within a self-consistency cycle for a single-band cuprate, we address a five-orbital molecular impurity using $20$ qubits and compute both its spectral function and self-energy for a fixed impurity model.
The spectral function is obtained directly on the real-frequency axis without analytic continuation, and the self-energy is computed through the Dyson equation for the impurity and bath orbitals together, which makes it causal by construction.
The same Lehmann representation also provides the Green's function at Matsubara frequencies, allowing the evaluation of $\Sigma(\ii\omega_n)$ for use in a DMFT self-consistency cycle.

The remainder of this paper is organized as follows.
Section~\ref{sec:method} describes the impurity model, the QSCI solver, and the Lehmann representation, explaining how the latter can be evaluated using eigenpairs obtained by different methods.
Section~\ref{sec:model} introduces the Fe-heme model and details the computational setup.
Section~\ref{sec:results} presents the ground-state accuracy, orbital occupations, the real-frequency spectral function, the self-energy, the charge gap, and the convergence of the classical post-processing.
Section~\ref{sec:discussion} discusses the implications for sampling-based impurity solvers and extensions to fault-tolerant quantum algorithms.
Finally, Sec.~\ref{sec:conclusion} summarizes our findings.
Additional computational details are provided in Appendix~\ref{app:conv}.

\section{Method}
\label{sec:method}

\subsection{DFT+DMFT and the Anderson impurity model}
\label{sec:method:dmft}

Dynamical mean-field theory maps an interacting lattice problem onto an effective Anderson impurity model (AIM), subject to a self-consistency condition~\cite{Metzner1989,Georges1996,Kotliar2006}.
In DFT+DMFT, the lattice Green's function is constructed from the Kohn--Sham Hamiltonian $\HKS(\mathbf{k})$ and the embedded impurity self-energy as
\begin{equation}
  G_{\mathrm{latt}}(\mathbf{k},\ii\omega_n)
  = \bigl[\,(\ii\omega_n + \chempot)I - \HKS(\mathbf{k})
          - \Sigma_{\mathrm{latt}}(\mathbf{k},\ii\omega_n)\,\bigr]^{-1},
  \label{eq:Glatt}
\end{equation}
where $\chempot$ is the chemical potential and $\omega_n$ is a
fermionic Matsubara frequency. We assume an orthonormal
single-particle basis and spin-independent one-body terms, and suppress spin indices in the Green's functions and self-energies.

Let $P(\mathbf{k})$ denote the rectangular projector from the
Kohn--Sham space onto the correlated orbital subspace.
The local Green's function is
\begin{equation}
G_{\mathrm{loc}}(\ii\omega_n) = \sum_{\mathbf{k}} P(\mathbf{k}) G_{\mathrm{latt}}(\mathbf{k},\ii\omega_n) P^{\dagger}(\mathbf{k}).
\label{eq:Gloc}
\end{equation}
Here and throughout, $\sum_{\mathbf{k}}$ denotes a normalized Brillouin-zone sum, including the appropriate integration weights.
The lattice self-energy is obtained by embedding the impurity
self-energy $\Sigma_{\mathrm{imp}}$ into the Kohn--Sham space:
\begin{equation}
    \Sigma_{\mathrm{latt}}(\mathbf{k},\ii\omega_n) = P^{\dagger}(\mathbf{k}) \left[
    \Sigma_{\mathrm{imp}}(\ii\omega_n) -V_{\rm DC}\right] P(\mathbf{k}),
    \label{eq:Slatt}
\end{equation}
where $V_{\rm DC}$ is the double-counting correction in the correlated subspace~\cite{CASTEPDMFT,DMFTforces}. It accounts approximately for interaction contributions already included in the DFT exchange-correlation functional.

The Weiss Green's function $\mathcal{G}_0$, which defines the noninteracting part of the effective impurity problem, is updated according to
\begin{equation}
\mathcal{G}_{0}^{-1}(\ii\omega_n) = G_{\mathrm{loc}}^{-1}(\ii\omega_n) + \Sigma_{\mathrm{imp}}(\ii\omega_n).
\label{eq:Weiss}
\end{equation}
The impurity one-body matrix is obtained from the local projection of the
Kohn--Sham Hamiltonian, with the double-counting correction
subtracted:
\begin{equation}
  \varepsilon^{\mathrm{imp}}
  =
  \sum_{\mathbf{k}}
  P(\mathbf{k})\HKS(\mathbf{k})P^{\dagger}(\mathbf{k})
  -V_{\mathrm{DC}}.
  \label{eq:epsimp}
\end{equation}
The hybridization function is then defined by
\begin{equation}
\Delta(\ii\omega_n)
=(\ii\omega_n+\mu)I-\varepsilon^\mathrm{imp}
-\mathcal{G}_{0}^{-1}(\ii\omega_n).
\label{eq:Delta}
\end{equation}

For a Hamiltonian-based impurity solver, the continuous bath is approximated by a finite set of bath orbitals. In a basis in which the bath one-body Hamiltonian is diagonal, the corresponding hybridization function is
\begin{equation}
[\Delta_\mathrm{AIM}(\ii\omega_n)]_{ab}
=
\sum_{\ell=0}^{\Nbath-1}
\frac{
    V_{a \ell}V_{b\ell}^{*}
  }{
    \ii\omega_n+\chempot-\epsbath{\ell}
  }.
 \label{eq:DeltaAIM}
\end{equation}
The bath energies $\epsbath{\ell}$ and hybridization amplitudes $V_{a\ell}$ are fitted to approximate $\Delta(\ii\omega_n)$ over a specified frequency range.
Together with the impurity one-body matrix and interaction
tensor, these parameters define the AIM Hamiltonian:
\begin{equation}
  \HAIM = \Himp + \Hbath + \Hhyb ,
  \label{eq:HAIM}
\end{equation}
with
\begin{align}
  \Himp
  &=
  \sum_{ab\sigma}
  \epsimp{ab}\,
  \dcr{a}{\sigma}\dan{b}{\sigma}
  \nonumber\\
  &\quad+
  \frac{1}{2}
  \sum_{\substack{abcd\\\sigma\sigma'}}
  U_{abcd}\,
  \dcr{a}{\sigma}\dcr{b}{\sigma'}
  \dan{d}{\sigma'}\dan{c}{\sigma},
  \label{eq:Himp}\\
 \Hbath
  &=
  \sum_{\ell\sigma}
  \epsbath{\ell}\,
  c_{\ell\sigma}^{\dagger}c_{\ell\sigma},
  \label{eq:Hbath}\\
  \Hhyb
  &=
  \sum_{a\ell\sigma}
  \left(
    V_{a\ell}\,
    \dcr{a}{\sigma}c_{\ell\sigma}
    +\mathrm{H.c.}
  \right).
  \label{eq:Hhyb}
\end{align}
Here $a,b,c,d$ label the $\Norb$ correlated impurity orbitals, $\ell$ labels the $\Nbath$ bath orbitals, and $\sigma,\sigma'$ label spin.
The operators $\dcr{a}{\sigma}$ and $c_{\ell\sigma}^{\dagger}$ create electrons in impurity and bath orbitals, respectively.
The interaction tensor $U_{abcd}$ denotes the unsymmetrized Coulomb matrix element $\langle ab|\hat{v}|cd\rangle$, consistent with the operator ordering in Eq.~\eqref{eq:Himp}.
The one-body energies do not include the chemical potential; the grand-canonical Hamiltonian is $\HAIM-\chempot\hat{N}$, where $\hat{N}$ is the total particle-number operator.

Solving the AIM yields the interacting impurity Green's
function $G_{\mathrm{imp}}$ and the self-energy
\begin{equation}
  \Sigma_{\mathrm{imp}}(\ii\omega_n)
  =
  \mathcal{G}_{0,\mathrm{AIM}}^{-1}(\ii\omega_n)
  -G_{\mathrm{imp}}^{-1}(\ii\omega_n),
  \label{eq:SigmaImp}
\end{equation}
where
\begin{equation}
  \mathcal{G}_{0,\mathrm{AIM}}^{-1}(\ii\omega_n)
  =
  (\ii\omega_n+\chempot)I
  -\varepsilon^{\mathrm{imp}}
  -\Delta_{\mathrm{AIM}}(\ii\omega_n).
  \label{eq:WeissAIM}
\end{equation}
The resulting self-energy is embedded through Eq.~\eqref{eq:Slatt} and used to update the lattice Green's function in Eq.~\eqref{eq:Glatt}.
The chemical potential is adjusted to reproduce the target total electron number, obtained by integrating the lattice spectral density with the Fermi--Dirac distribution.
The lattice and impurity problems use the same chemical potential; the impurity occupation is determined by the resulting bath and need not equal its nominal atomic value.
The procedure is repeated until self-consistency is reached.
In the limit of an exact bath representation and an exact impurity solution, the converged solution satisfies $G_{\mathrm{imp}}=G_{\mathrm{loc}}$.

For a fixed correlated subspace and interaction tensor, the number of impurity and bath orbitals and the form of the interaction remain unchanged throughout the DMFT loop.
The bath energies and hybridization amplitudes are updated to represent the evolving hybridization function.
The impurity one-body matrix may also change if the double-counting correction or the underlying Kohn--Sham Hamiltonian is updated.
The impurity solver therefore acts as a repeatedly invoked subroutine with a fixed operator structure, making it a natural target for quantum acceleration.

In this work, we adopt a \emph{stepwise} hybrid workflow.
The AIM is constructed through a classical DFT+DMFT calculation on Fugaku and subsequently solved using QSCI, with quantum sampling performed on Quantinuum Reimei and classical post-processing.
All impurity-model parameters and the chemical potential are held fixed during this stage, and the QSCI results are not fed back into the DFT+DMFT loop.
This workflow allows us to assess the accuracy of the QSCI impurity solver independently of the propagation of solver errors through DMFT self-consistency.

\subsection{QSCI as an impurity solver}
\label{sec:method:qsci}

QSCI~\cite{Kanno2023} uses quantum
measurements to select a set of Slater determinants for
classical subspace diagonalization.
It builds on selected configuration interaction
methods~\cite{Huron1973,Holmes2016,Tubman2016}, including
their application to impurity problems through truncated
CI expansions~\cite{Zgid2012}.
In QSCI, determinants are selected by sampling a quantum
state rather than by a classical selection criterion.

A trial state $\trial$ is prepared on $\Nq = 2(\Norb + \Nbath)$ qubits using the Jordan--Wigner mapping and measured in the computational basis.  Each measurement outcome is a bit string specifying the occupations of the spin orbitals and hence a Slater determinant $\conf{\phi}$.
After discarding outcomes outside the target symmetry sector $\HilN$, we collect the distinct observed determinants into the set
\begin{equation}
  \SN = \bigl\{\, \conf{\phi} \in \HilN \;:\;
        \text{observed at least once} \,\bigr\}.
  \label{eq:SN}
\end{equation}
The AIM Hamiltonian $\HAIM$ is then projected onto $\spn{\SN}$ and diagonalized classically to obtain the approximate ground state $\GSq$ and its energy $\tilde{E}_{0}^{N}$.

Finite sampling and device noise affect which determinants are selected, but the Hamiltonian matrix elements within the selected subspace are evaluated classically.
Consequently, for any nonempty selected subspace contained in $\HilN$, the lowest eigenvalue satisfies the variational bound
\begin{equation}
  \tilde{E}_{0}^{N}\ge E_{0}^{N},
  \label{eq:variational}
\end{equation}
where $E_{0}^{N}$ is the exact ground-state energy in the target sector.

The AIM Hamiltonian conserves the numbers of spin-up and
spin-down electrons separately, so its eigenstates can be
classified by $(N_{\uparrow},N_{\downarrow})$.
We therefore construct each subspace within a fixed
spin-resolved particle-number sector.

\subsection{Lehmann representation of the impurity Green's function}
\label{sec:method:lehmann}

At zero temperature, the retarded impurity Green's function is expressed through the Lehmann representation as
\begin{align}
  \Gret{ab,\sigma}(\omega + \ii\broad)
  ={}& \sum_{m}
    \frac{\langle \Psi_{0}^{N} \rvert \dan{a}{\sigma} \exNp{m}
          \langle \Psi_{m}^{N+1} \rvert \dcr{b}{\sigma} \GSex}
         {\omega + \chempot - (\ENp{m} - \EGS) + \ii\broad}
  \nonumber\\[2pt]
  &+ \sum_{n}
    \frac{\langle \Psi_{0}^{N} \rvert \dcr{b}{\sigma} \exNm{n}
          \langle \Psi_{n}^{N-1} \rvert \dan{a}{\sigma} \GSex}
         {\omega + \chempot + (\ENm{n} - \EGS) + \ii\broad}.
  \label{eq:lehmann}
\end{align}
Here $|\Psi_0^N\rangle$ is an $N$-electron ground state, and $|\Psi_m^{N+1}\rangle$ and $|\Psi_n^{N-1}\rangle$ are eigenstates in the particle-addition and particle-removal sectors, respectively.
All energies are eigenvalues of $\HAIM$, without the chemical-potential term.
The frequency $\omega$ is measured relative to $\chempot$, and $\broad>0$ is a finite broadening parameter; the limit $\broad\to0^+$ gives the retarded boundary value.
For a ground state with fixed $(N_{\uparrow},N_{\downarrow})$, the sums run over the sectors reached by adding or removing an electron with spin $\sigma$.

The spin-resolved orbital spectral function is
\begin{equation}
  A_{a\sigma}(\omega)
  =
  -\frac{1}{\pi}
  \operatorname{Im}
  G^{R}_{aa,\sigma}(\omega+\ii\broad).
  \label{eq:spectral}
\end{equation}
The spin-summed orbital and total impurity spectral functions are defined as $A_a(\omega)=\sum_{\sigma}A_{a\sigma}(\omega)$ and $A_{\mathrm{tot}}(\omega)=\sum_a A_a(\omega)$, respectively.
The diagonal Lehmann weights are
\begin{align}
  \wplus{m}{a\sigma}
  &=
  \left|
    \langle \Psi_m^{N+1} |
    \dcr{a}{\sigma}
    | \Psi_0^N \rangle
  \right|^2,
  \nonumber\\
  \wminus{n}{a\sigma}
  &=
  \left|
    \langle \Psi_n^{N-1} |
    \dan{a}{\sigma}
    | \Psi_0^N \rangle
  \right|^2.
  \label{eq:weights}
\end{align}

The Lehmann representation does not depend on the method used to obtain the eigenstates and eigenvalues.
Using complete eigenstate sets gives the exact Green's function of the finite AIM, whereas using approximate eigenpairs in restricted subspaces yields an approximate Green's function whose accuracy depends on those subspaces and eigenpairs.

Equation~\eqref{eq:lehmann} requires the ground state in the $N$-electron sector, the energies of the relevant states in the $N\pm1$ sectors, and the transition matrix elements connecting these states.
Extensions based on quantum phase estimation (QPE)~\cite{Kitaev1995,Abrams1999,AspuruGuzik2005} or quantum singular value transformation (QSVT)~\cite{Gilyen2019,Low2019} may provide alternative routes to the spectral information, although the quantum algorithms and their interface with classical post-processing differ from those used in QSCI, as discussed in Sec.~\ref{sec:disc:ft}.

In the present QSCI implementation, each determinant is
stored as a pair of bit strings specifying the spin-up
and spin-down occupations.
The action of $\dan{a}{\sigma}$ or $\dcr{a}{\sigma}$,
including the fermionic sign associated with the chosen
Jordan--Wigner ordering, is evaluated using bitwise
operations.
For a fixed orbital and spin, the particle-removal and
particle-addition vectors,
$\dan{a}{\sigma}\GSq$ and
$\dcr{a}{\sigma}\GSq$, can therefore be constructed
in a single pass over the ground-state determinant basis,
with a computational cost linear in its size.
The Lehmann weights in Eq.~\eqref{eq:weights} are then
obtained from overlaps with the corresponding
$N\pm1$ eigenstates, without explicitly constructing
matrix representations of the creation or annihilation
operators.

The operators $\dan{a}{\sigma}$ and $\dcr{a}{\sigma}$ change $N_{\sigma}$ by $-1$ and $+1$, respectively, while leaving the opposite-spin particle number unchanged.
Thus, only the particle-number sectors adjacent to $(N_{\uparrow},N_{\downarrow})$ contribute to Eq.~\eqref{eq:lehmann}.

Two exact relations provide internal consistency checks.
The first is the spectral-weight sum rule,
\begin{equation}
  \sum_{m} \wplus{m}{a\sigma} + \sum_{n} \wminus{n}{a\sigma} = 1 ,
  \label{eq:sumrule}
\end{equation}
which follows from the canonical anticommutation relations $\{\hat{d}_{a\sigma},\hat{d}^\dagger_{b\sigma'}\}=\delta_{ab}\delta_{\sigma\sigma'}$.
In the QSCI calculation, this identity is preserved by construction: the augmentation rule in Eq.~\eqref{eq:aug} ensures that both $\dan{a}{\sigma}\GSq$ and $\dcr{a}{\sigma}\GSq$ lie entirely within the corresponding $\spn{\SNpm}$ subspaces.
Provided that complete orthonormal eigenstate sets are
used within these subspaces, Eq.~\eqref{eq:sumrule}
therefore checks the implementation and numerical
completeness of the spectral calculation rather than
the accuracy of the sampled ground state.

The second relation connects the orbital occupation to the particle-removal weights:
\begin{equation}
  \langle \nop{a}{\sigma} \rangle
  =
  \sum_n \wminus{n}{a\sigma}
  =
  \lim_{\broad\to0^+}
  \int_{-\infty}^{0}
  \Aspec{a\sigma}(\omega)\,\dd\omega.
  \label{eq:occ}
\end{equation}
The last equality assumes that all particle-removal poles lie below zero frequency and all particle-addition poles lie above it.
For the exact eigenenergies, this requires
\begin{equation}
  E_0^N-E_0^{N-1}
  <
  \chempot
  <
  E_0^{N+1}-E_0^N,
  \label{eq:muwindow}
\end{equation}
where $E_0^{N\pm1}$ denotes the lowest energy among the relevant particle-addition or particle-removal sectors.
For approximate spectra, the corresponding condition must hold for the approximate energies used in the Lehmann representation.
For the fixed Hamiltonian and particle-number sectors used here, changing $\chempot$ only shifts the frequency origin (Sec.~\ref{sec:model:aim}).
At finite $\broad$, the negative-frequency integral receives contributions from the broadened tails of both particle-addition and particle-removal poles, whereas the sum of particle-removal weights remains equal to the orbital occupation.

The quantum device samples only the target $N$-electron sector.
The determinant sets for the $N+1$ and $N-1$ sectors are constructed classically by applying single-electron creation and annihilation operators, respectively, to each sampled determinant:
\begin{align}
  \SNp
  &=
  \bigl\{
    \hat{a}^{\dagger}_{p\sigma}\conf{\phi}
    \;:\;
    \conf{\phi}\in\SN,\
    \hat{a}^{\dagger}_{p\sigma}\conf{\phi}\ne0
  \bigr\},
  \nonumber\\
  \SNm
  &=
  \bigl\{
    \hat{a}^{\phantom{\dagger}}_{p\sigma}\conf{\phi}
    \;:\;
    \conf{\phi}\in\SN,\
    \hat{a}^{\phantom{\dagger}}_{p\sigma}\conf{\phi}\ne0
  \bigr\}.
  \label{eq:aug}
\end{align}%
Here $\hat{a}^{\dagger}_{p\sigma}$ creates an electron in the $p$-th spatial orbital with spin $\sigma$, with $p$ running over all impurity and bath orbitals, so that $\hat{a}^{\dagger}_{p\sigma}$ denotes $\dcr{a}{\sigma}$ for impurity orbitals and $c^{\dagger}_{\ell\sigma}$ for bath orbitals.  Both spin components are included.
The resulting configurations are stored in a canonical determinant convention, and duplicates are removed.
The Hamiltonian is then diagonalized within the span of each augmented set, separately for each spin-resolved particle-number sector.

Particle addition and removal generate subspaces of different sizes and coverage.
Each sampled determinant contains twelve occupied and eight unoccupied spin orbitals, allowing twelve particle-removal operations and eight particle-addition operations.
For the $444$ sampled determinants, which cover $1.0\%$ of the target $N$-electron sector, the resulting addition and removal subspaces cover $4.2\%$ and $3.3\%$ of their respective combined sectors (Table~\ref{tab:summary}).
Thus, although particle removal generates more distinct configurations, they span a smaller fraction of the available sector space.
The implications of this asymmetry for the spectral accuracy are examined in Sec.~\ref{sec:results:spectral}.

\section{Model and computational setup}
\label{sec:model}

\subsection{Heme $b$ and its correlated electronic structure}
\label{sec:model:system}

Heme $b$, an iron protoporphyrin~IX complex, is the prosthetic group of hemoglobin, myoglobin, and many cytochromes.
Its iron center supports several low-lying spin states, whose relative energies are sensitive to coordination and electronic correlations.
Experimental studies of four-coordinate ferrous porphyrins have identified an intermediate-spin $S=1$ ground state~\cite{Mispelter1980,Boyd1979}, providing an important benchmark for electronic-structure methods.

Previous DFT+DMFT studies have investigated the role of local electronic correlations in heme complexes~\cite{Weber2013,Weber2014} and other transition-metal complexes~\cite{Chen2015}.
Here we retain the full five-orbital iron $3d$ shell as the correlated subspace, with $\Norb=5$, to describe the interplay of orbital occupations, spin states, and local correlations.
Compared with models restricted to a subset of the  $d$ orbitals, this choice places greater demands on the impurity solver while retaining all local correlation channels within the iron $3d$ shell.

\subsection{Molecular geometry and DFT calculations}
\label{sec:model:dft}

We consider the complete heme $b$ molecule, C$_{34}$H$_{32}$FeN$_4$O$_4$, including the porphyrin macrocycle and its vinyl, methyl, and propionate substituents.
The initial geometry was taken from Ref.~\cite{Weber2013} and reoptimized using the computational setup adopted here.
Whereas the earlier study used a molecular code, we employ periodic boundary conditions, with the molecule placed in a supercell providing at least $4.8$~\AA{} of vacuum in the macrocycle plane.
The DFT calculations were performed using GPAW~\cite{GPAW2010}, the PBE exchange-correlation functional~\cite{PBE1996}, and a double-zeta polarized numerical atomic-orbital basis.
Brillouin-zone sampling was restricted to the $\Gamma$ point.
A spin-unpolarized calculation provides the Kohn--Sham reference from which the impurity model is constructed.

The optimized structure has a four-coordinate iron center with no axial ligand.
The Fe--N bond lengths are approximately $1.96$~\AA{}, and the iron atom lies within $0.001$~\AA{} of the plane defined by the four coordinating nitrogen atoms.
The local FeN$_4$ coordination therefore approximates a square-planar geometry.
We label the iron $3d$ basis orbitals in the order
\begin{equation}
  a=0,1,2,3,4
  \;\longleftrightarrow\;
  d_{x^{2}-y^{2}},\;
  d_{yz},\;
  d_{z^{2}},\;
  d_{xz},\;
  d_{xy}.
  \label{eq:orbitalorder}
\end{equation}
These labels refer to real $d$ harmonics defined in a frame whose $z$ axis is normal to the FeN$_4$ plane and whose $x$ and $y$ axes bisect the Fe--N bonds.
We use these orbital labels throughout, while noting that deviations from ideal $D_{4h}$ symmetry prevent their identification with exact symmetry-adapted orbitals.

\subsection{The Anderson impurity model and parameters}
\label{sec:model:aim}

The correlated subspace consists of the five localized  Fe $3d$ orbitals, coupled to $\Nbath=5$ bath orbitals obtained by discretizing the hybridization function.
The resulting AIM contains $\Norb+\Nbath=10$ spatial orbitals and is encoded on $\Nq=20$ qubits using the Jordan--Wigner transformation.
We target the $N=12$ electron sector with $(N_{\uparrow},N_{\downarrow})=(6,6)$.
The ground state is computed in this sector, while the adjacent particle-addition and particle-removal sectors enter the Green's function calculation.
All results reported below are evaluated at zero temperature.

The diagonal one-body parameters are listed in Table~\ref{tab:aimparams}.
The impurity levels include a double-counting correction~\cite{Czyzyk1994} and lie substantially below the bath energies.
Their relation to the interacting spectral features is discussed in Sec.~\ref{sec:results:sigma}.
The impurity one-body matrix is nearly diagonal and the bath Hamiltonian is diagonal.
The hybridization amplitudes $V_{a\ell}$ form a real $5\times5$ matrix; its largest elements are given in Appendix~\ref{app:conv}.

We retain the full rotationally invariant interaction tensor $U_{abcd}$ of Eq.~\eqref{eq:Himp}, including the terms with three or four distinct orbital indices that are absent from the Kanamori form~\cite{Kanamori1963,Georges2013}.
The interaction parameters follow the Slater-integral parametrization used in Ref.~\cite{Weber2013}, with $F^0=4$~eV and a Hund's coupling parameter $J=0.8$~eV.
The intra-orbital repulsion is identical for all five orbitals, $U_{aa}=F^0+\tfrac{8}{7}J\simeq4.914$~eV.
The pair-dependent exchange integrals $J_{ab}$ range from $0.34$ to $0.71$~eV, and the interorbital repulsions satisfy $U'_{ab}=U_{aa}-2J_{ab}$.
Restricting the interaction to the Kanamori form raises the ground-state energy by $0.40$~eV (Appendix~\ref{app:conv}).

Unless stated otherwise, we use a broadening of $\broad=0.3$~eV in Eq.~\eqref{eq:lehmann}.
The one-body energies of the AIM are measured from the chemical potential of the DFT+DMFT calculation, and we use $\chempot=0$ unless stated otherwise.
Following the paramagnetic DMFT solution, we take the ground state from the $S_z=0$ sectors, among which the $(6,6)$ sector has the lowest energy.
For this ground state and the adjacent sectors $(5,6)$, $(6,5)$, $(7,6)$, and $(6,7)$, the stability interval of Eq.~\eqref{eq:muwindow} is $-2.131<\chempot<-0.023$~eV.
The DMFT chemical potential therefore lies $23$~meV above its upper bound, and the lowest particle-addition pole appears at $\omega=-0.023$~eV, just below zero frequency.
In the real-frequency figures and in Sec.~\ref{sec:results:gap}, we set $\chempot=-0.70$~eV, inside this interval, so that zero frequency lies within the gaps of both the impurity spectrum and the DFT+DMFT DOS.
Throughout this work, particle-addition and particle-removal contributions are identified by their Lehmann sectors rather than by the sign of frequency alone.

\begin{table}[t]
\caption{\label{tab:aimparams}%
Diagonal one-body parameters of the Fe-heme
Anderson impurity model, in eV.
The impurity entries are
$\varepsilon^{\mathrm{imp}}_{aa}$ and include
the double-counting correction.
Off-diagonal impurity matrix elements have
magnitudes of at most $0.10$~eV.
Bath energies are given in the diagonal bath basis;
bath indices do not imply a one-to-one
correspondence with impurity orbitals.
Both spin components share identical parameters.
}
\begin{ruledtabular}
\setlength{\tabcolsep}{3pt}
\begin{tabular}{lccccc}
$a$
& $d_{x^{2}-y^{2}}$ & $d_{yz}$ & $d_{z^{2}}$
& $d_{xz}$ & $d_{xy}$ \\
$\varepsilon^{\mathrm{imp}}_{aa}$
& $-25.810$ & $-25.445$ & $-25.820$
& $-25.466$ & $-26.384$ \\
\colrule
$\ell$
& $0$ & $1$ & $2$ & $3$ & $4$ \\
$\epsbath{\ell}$
& $-16.171$ & $-6.402$ & $-1.915$
& $+0.576$ & $+5.489$ \\
\end{tabular}
\end{ruledtabular}
\end{table}

\subsection{Hartree--Fock trial-state preparation}
\label{sec:model:trial}

The trial state used for quantum sampling is the broken-symmetry unrestricted Hartree--Fock (HF) solution of the AIM Hamiltonian in the $(6,6)$ sector, obtained classically.
The spin-restricted HF determinant is a spin singlet, whereas the ground state in this sector is the $S_z=0$ component of a spin triplet (Sec.~\ref{sec:results:gs}); its overlap with the ground state therefore vanishes.
In the unrestricted solution, the spin-up and spin-down orbitals differ while $N_{\uparrow}=N_{\downarrow}$, and the resulting determinant has a squared overlap of $0.44$ with the exact ground state.
We prepare this state using a real orbital rotation~\cite{Kivlichian2018}, which represents the transformation from the original impurity and bath orbitals to the canonical HF orbitals as a sequence of Givens rotations.

Let $\hat{a}_i^{\dagger}$ create an electron in an original spin orbital, with $i$ running over both impurity and bath spin orbitals.
We define the canonical HF creation operators by
\begin{equation}
  \hat{b}_p^{\dagger}
  =
  \sum_i R_{ip}\hat{a}_i^{\dagger},
  \label{eq:HFrotation}
\end{equation}
where $R$ is a real orthogonal orbital-transformation matrix.
Starting from a reference occupation state $|\mathrm{Ref}\rangle$ with the target $(N_{\uparrow},N_{\downarrow})$, the HF determinant is prepared as
\begin{equation}
  |\Psi_{\mathrm{HF}}\rangle
  =
  \exp\left[
    \sum_{ij}\kappa_{ij}
    \hat{a}_i^{\dagger}\hat{a}_j
  \right]
  |\mathrm{Ref}\rangle,
  \label{eq:HFstate}
\end{equation}
where $\kappa$ is real and antisymmetric, $\kappa^{T}=-\kappa$, and satisfies $R=e^{\kappa}$.
The rotation acts separately within the spin-up and spin-down orbital spaces and therefore preserves both $N_{\uparrow}$ and $N_{\downarrow}$; for the unrestricted solution, the two spin blocks of $R$ differ.
The phases of the canonical orbitals can be chosen such that each spin block of $R$ has determinant $+1$, allowing a real antisymmetric generator.

This construction is consistent with the orbital-rotation description of Slater determinants underlying Thouless' theorem~\cite{Thouless1960}.
The orbital transformation is decomposed into Givens rotations to obtain a particle-number-conserving circuit for preparing the HF determinant; only rotations between occupied and unoccupied orbitals are required, $N_{\sigma}(\Norb+\Nbath-N_{\sigma})=24$ for each spin.

\subsection{Sampling on the quantum processor}
\label{sec:model:sampling}

The trial state is prepared and measured using $\Nq=20$ of the $56$ qubits of Quantinuum Reimei, a System Model~H2 trapped-ion quantum computer installed at the RIKEN Wako campus~\cite{QuantinuumRIKENH22026}.
The compiled state-preparation circuit contains $96$ two-qubit gates.
The measurements reported here were performed in October 2026.

Computational-basis measurements yield bit strings specifying the spin-orbital occupations.
Because the ideal state-preparation circuit preserves $N_{\uparrow}$ and $N_{\downarrow}$ separately, outcomes outside the target $(N_{\uparrow},N_{\downarrow})=(6,6)$ sector are attributed to device errors and discarded.
The $30\,000$ shots yield $931$ distinct bit strings before this postselection.
We then augment the retained determinant set with the configurations obtained by exchanging the spin-up and spin-down occupation strings, making the set closed under time reversal up to overall phase factors.
The resulting set contains $|\SN|=444$ determinants.
No minimum occurrence count is imposed: every determinant observed in the target sector is retained.

\subsection{Classical post-processing}
\label{sec:model:classical}

All classical post-processing is performed on a conventional workstation.
The ED reference is obtained with the Lanczos method in the full sectors.
The AIM Hamiltonian is first diagonalized within
$\spn{\SN}$ to obtain the approximate ground state
$\GSq$.
The particle-addition and particle-removal determinant sets $\SNpm$ are then constructed using the augmentation rule in Eq.~\eqref{eq:aug}, and the Hamiltonian is diagonalized within their spans, separately for each spin-resolved particle-number sector.
The Green's function and spectral function are assembled using Eq.~\eqref{eq:lehmann}, and the self-energy is obtained from Eq.~\eqref{eq:SigmaFull} (Sec.~\ref{sec:results:sigma}).

Two numerical truncations can be introduced at this stage: a limit $\Mkeep$ on the number of eigenstates retained in each adjacent sector, and a cap on the number of determinants in $\SNm$, which is larger than $\SNp$ in the present calculation.
Both truncations affect only the classical post-processing and can be varied using the same measurement data, without additional quantum-device executions.
Unless otherwise stated, the results below retain all eigenstates within each augmented subspace and impose no additional cap on the determinant sets.
Appendix~\ref{app:conv} examines the effects of both truncations.

\section{Results}
\label{sec:results}

\subsection{Ground-state accuracy and convergence}
\label{sec:results:gs}

Table~\ref{tab:summary} summarizes the main results discussed in this section.
Quantum sampling on Quantinuum Reimei, followed by postselection and augmentation under time reversal, yields $|\SN|=444$ determinants, corresponding to $1.0\%$ of the $44\,100$-dimensional target sector.
Diagonalizing $\HAIM$ within $\spn{\SN}$ gives a ground-state energy of $-158.9906$~eV, compared with the full-sector ED value of $-159.0391$~eV.
The QSCI energy lies approximately $49$~meV above the ED result, consistent with the variational bound.

\begin{table*}[t]
\caption{\label{tab:summary}%
Summary of the calculation.  ``ED'' denotes exact diagonalization within the full target sectors. ``QSCI'' denotes calculations using the $444$-determinant ground-state subspace obtained from measurements on Quantinuum Reimei, postselection, and augmentation under time reversal, together with the associated augmented particle-addition and particle-removal subspaces.
Orbital occupations are quoted per spin, as in Fig.~\ref{fig:gs}(c).
The final column reports signed QSCI-minus-ED differences for energies and occupations, the infidelity $1-F$, sector coverage for subspace dimensions, and the spectral errors defined in Eq.~\eqref{eq:spectral_error}.
The $N+1$ and $N-1$ dimensions are summed over the two relevant spin-resolved sectors.
The final row reports the Lehmann sum rule, Eq.~\eqref{eq:sumrule}.
}
\begin{ruledtabular}
\begin{tabular}{lccc}
Quantity & ED & QSCI & Deviation / diagnostic \\
\colrule
Ground-state energy (eV)             & $-159.0391$ & $-158.9906$ & $49$~meV \\
Fidelity $F$                         & $1$ & $0.9825$ & $1.75\times10^{-2}$ \\
Subspace dimension, $N$              & $44\,100$ & $444$ & $1.0\%$ of sector \\
\quad $N+1$ (addition)          & $50\,400$ & $2138$ & $4.2\%$ of sector \\
\quad $N-1$ (removal)           & $105\,840$ & $3442$ & $3.3\%$ of sector \\[3pt]
$\langle n_{a\sigma}\rangle$, $d_{x^{2}-y^{2}}$       & $0.9836$ & $0.9917$ & $+8.1\times10^{-3}$ \\
\quad $d_{yz}$                       & $0.5066$ & $0.5001$ & $-6.5\times10^{-3}$ \\
\quad $d_{z^{2}}$                    & $0.9633$ & $0.9836$ & $+2.03\times10^{-2}$ \\
\quad $d_{xz}$                       & $0.5318$ & $0.5130$ & $-1.88\times10^{-2}$ \\
\quad $d_{xy}$              & $0.5361$ & $0.5352$ & $-8.5\times10^{-4}$ \\
Total impurity occupancy $n_{d}$ (both spins) & $7.043$ & $7.047$ & $+4.4\times10^{-3}$ \\[3pt]
Spectral error $\epsilon_A$ (eV$^{-1}$), total                   & --- & --- & $3.5\times10^{-2}$ \\
\quad removal contribution                & --- & --- & $2.95\times10^{-2}$ \\
\quad addition contribution               & --- & --- & $1.85\times10^{-2}$ \\
Lehmann sum rule                     & $1.0000000$ & $1.0000000$ & $2\times10^{-15}$ \\[3pt]
\end{tabular}
\end{ruledtabular}
\end{table*}

Energy agreement alone does not fully characterize the accuracy of the ground state. For a normalized state close to an exact eigenstate, the energy error is second order in the wave-function error.
We therefore also evaluate the ground-state fidelity,
\begin{equation}
  F
  =
  \bigl|
    \langle \tilde{\Psi}_0^N | \Psi_0^N \rangle
  \bigr|^2
  =
  0.9825,
  \label{eq:fidelity}
\end{equation}
corresponding to an infidelity $1-F=1.75\times10^{-2}$.
The accuracy of the dynamical quantities also depends on the particle-addition and particle-removal subspaces, as examined in Secs.~\ref{sec:results:spectral} and~\ref{sec:results:sigma}.

To assess spectral convergence, we define the root-mean-square spectral error as
\begin{equation}
  \epsilon_A
  =
  \sqrt{
    \frac{1}{\omega_{\max}-\omega_{\min}}
    \int_{\omega_{\min}}^{\omega_{\max}}
    \bigl[
      A_{\mathrm{QSCI}}(\omega)
      -A_{\mathrm{ED}}(\omega)
    \bigr]^2
    \,\mathrm{d}\omega
  },
  \label{eq:spectral_error}
\end{equation}%
where both spectra are evaluated using the same broadening $\eta=0.3$~eV and chemical potential, and the interval $[\omega_{\min},\omega_{\max}]=[-20,6]$~eV is held fixed throughout.  Normalizing by the interval length gives $\epsilon_A$ the units of the spectral function, eV$^{-1}$, so that it can be compared directly with the peak height of $A_{\mathrm{tot}}(\omega)$, about $2.5$~eV$^{-1}$.
The corresponding errors for the particle-removal and particle-addition contributions are defined analogously using their respective spectra.

\begin{figure*}[t]
\centering
\includegraphics[width=\textwidth]{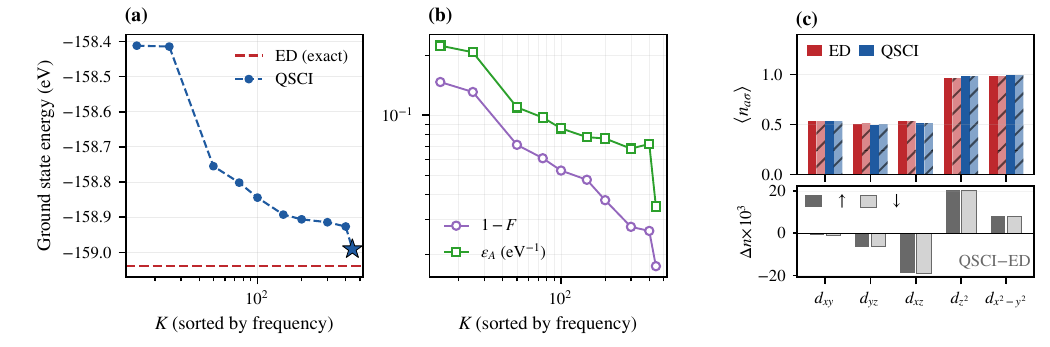}
\caption{\label{fig:gs}%
Ground-state accuracy and convergence for the Fe-heme impurity model.
(a) Ground-state energy as a function of the number of retained determinants, $K$, ordered by measurement count, with the time-reversal partner of each determinant placed immediately after it.
The dashed line denotes the full-sector ED reference, and the star marks the complete QSCI subspace, $|\SN|=444$, including configurations added under time reversal.
(b) Infidelity $1-F$ and spectral error $\epsilon_A$, defined in Eq.~\eqref{eq:spectral_error}, as functions of the number of retained determinants.
(c) Spin-resolved impurity occupations from ED and QSCI.
The lower panel shows the differences, QSCI minus ED, on an expanded scale.
Solid and hatched bars distinguish the two spin components.
}
\end{figure*}

Figure~\ref{fig:gs}(a) shows the convergence
of the ground-state energy with the number
of retained determinants.
Figure~\ref{fig:gs}(b) shows the corresponding
infidelity and spectral error.
The infidelity decreases monotonically and the spectral error nearly so over the range considered, with no apparent saturation.
The last determinants added, which were observed only once or were added as their time-reversal partners, still reduce both errors substantially.

Among the $200$ determinants with the largest weights in the exact ground state, $28$ are absent from $\SN$; they carry a combined weight of $3.7\times10^{-3}$.
The total ground-state weight outside $\spn{\SN}$ is $5.6\times10^{-3}$, so no state in $\spn{\SN}$ can reach a fidelity above $0.994$.
The fidelity obtained is lower, $0.9825$, because the absent determinants also affect the coefficients of the retained ones in the subspace diagonalization.
None of these missing determinants was observed in the measurement data.
The result therefore points to incomplete sampling of relevant configurations as the principal limitation of the present ground-state calculation.

The ED ground state is a spin triplet, with $\langle\hat{S}^{2}\rangle=2$ to numerical precision, where $\hat{S}$ is the total spin of the impurity and bath combined and $\hbar=1$; the $(6,6)$ sector contains its $S_z=0$ component.
This agrees with the intermediate-spin $S=1$ ground state identified experimentally for four-coordinate ferrous porphyrins (Sec.~\ref{sec:model:system}).
Two further triplets lie only $45$ and $100$~meV above the ground state, and the lowest singlet lies $0.985$~eV above it.
The QSCI ground state is also a triplet.
However, QSCI places its lowest excited triplet at $138$~meV and its lowest singlet $0.17$~eV above the ED value.
Thus, although the calculation reproduces the triplet ground state, it does not yet reproduce the low-lying excitation energies quantitatively.
These closely spaced spin states provide a more stringent test of the impurity solver than the ground-state energy alone.

\subsection{Orbital occupations}
\label{sec:results:occ}

The spin-resolved impurity occupations are shown in Fig.~\ref{fig:gs}(c).
Two orbitals, $d_{x^{2}-y^{2}}$ and $d_{z^{2}}$, are nearly doubly occupied, with $\langle n_{a\sigma}\rangle=0.98$ and $0.96$, whereas the other three, $d_{yz}$, $d_{xz}$, and $d_{xy}$, are close to half filling, with $\langle n_{a\sigma}\rangle=0.51$--$0.54$.
The $d_{xz}$ and $d_{yz}$ occupations are similar, consistent with their symmetry equivalence in an ideal $D_{4h}$ environment, although the actual molecular structure and model parameters depart from this ideal symmetry.
The two spin components have equal orbital occupations, $\langle\nop{a}{\uparrow}\rangle =\langle\nop{a}{\downarrow}\rangle$, as expected for the $S_z=0$ component of the triplet ground state.
Both the Hamiltonian and the target $(N_{\uparrow},N_{\downarrow})=(6,6)$ sector are invariant under spin reversal.

QSCI reproduces each spin-resolved occupation to within $2.1\times10^{-2}$ electrons of ED.
The largest deviations occur for $d_{z^{2}}$ and $d_{xz}$, with opposite signs.
The total impurity occupation is $n_d\simeq7.04$ electrons in ED, compared with $7.05$ in QSCI.
This is larger than the range $6.2$--$6.8$ reported in Ref.~\cite{Weber2013}.
The correlated-orbital occupation depends on the definition of the localized subspace and on the impurity-model construction, so these values need not coincide across calculations.
The occupations evaluated directly from the ground state agree, to numerical precision, with the sums of the particle-removal Lehmann weights in Eq.~\eqref{eq:occ}.

\subsection{Real-frequency spectral function}
\label{sec:results:spectral}

\begin{figure*}[t]
\centering
\includegraphics[width=\textwidth]{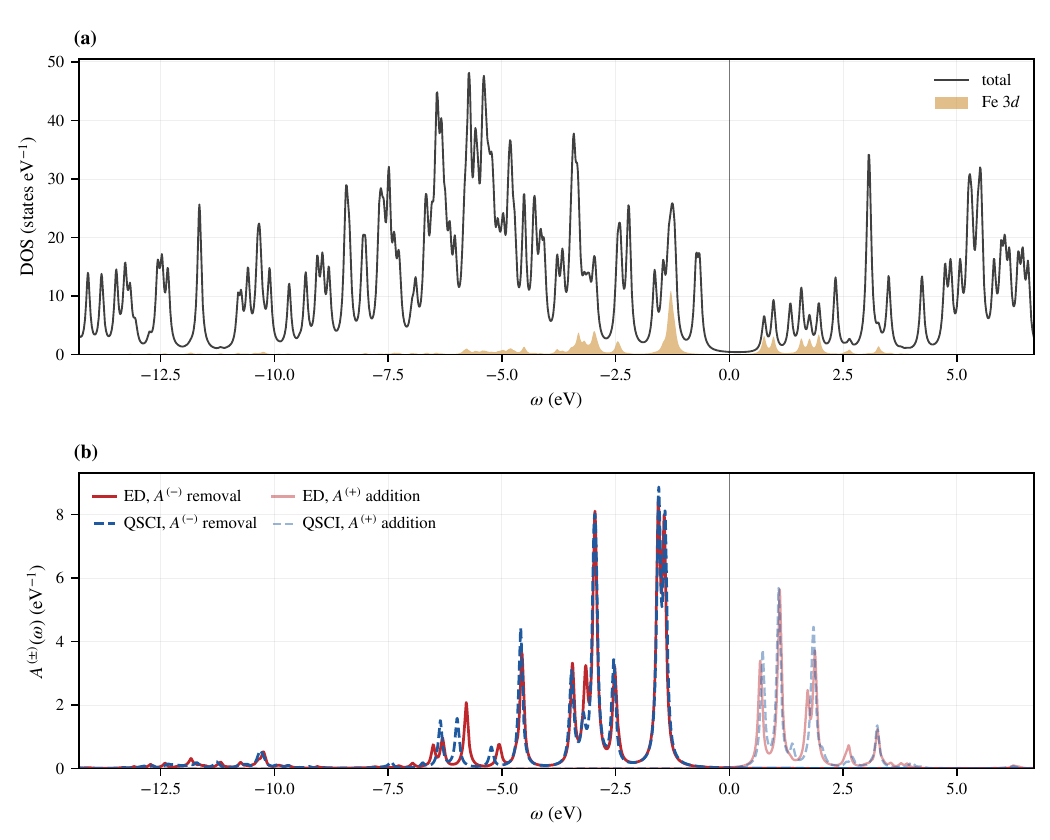}
\caption{\label{fig:spectral}%
Impurity spectral function on the real-frequency axis, obtained from the Lehmann representation without analytic continuation, and the density of states (DOS) of the DFT+DMFT calculation.
Frequencies are measured relative to the chemical potential.
(a) Total DOS and its Fe $3d$ contribution, with a broadening of $0.05$~eV.
(b) Particle-removal and particle-addition contributions, $\Aminus{}(\omega)$ and $\Aplus{}(\omega)$, respectively, separated according to their Lehmann terms, obtained using exact diagonalization (ED) and QSCI.
Both calculations use a broadening of $\eta=0.05$~eV.
}
\end{figure*}

Figure~\ref{fig:spectral}(b) shows the impurity spectral function obtained from the Lehmann representation, Eq.~\eqref{eq:lehmann}, using the QSCI ground state and the classically constructed particle-addition and particle-removal subspaces.
The spectral function is evaluated directly at real frequencies without analytic continuation.
Figure~\ref{fig:spectral}(a) shows the density of states (DOS) of the DFT+DMFT calculation from which the AIM is constructed, together with its Fe $3d$ contribution.

The QSCI spectrum is in good quantitative agreement with the ED reference over the frequency range shown.
The spectral error defined in Eq.~\eqref{eq:spectral_error} is
\begin{equation}
  \epsilon_A
  =
  3.5\times10^{-2}~\mathrm{eV}^{-1}.
  \label{eq:spectral_error_total}
\end{equation}
The calculation reproduces the main spectral features, including the dominant peak positions and weights.

The Lehmann representation separates the particle-addition and particle-removal contributions, allowing the accuracy of each to be assessed independently.
Figure~\ref{fig:spectral}(b) shows $\Aplus{}(\omega)$ and $\Aminus{}(\omega)$, classified by their respective Lehmann terms.
The spectral error of the removal contribution is approximately $1.6$ times that of the addition contribution (Table~\ref{tab:summary}).

This asymmetry is consistent with the smaller coverage of the removal sectors (Sec.~\ref{sec:method:lehmann} and Table~\ref{tab:summary}).

Both QSCI and ED satisfy the Lehmann sum rule, Eq.~\eqref{eq:sumrule}, to numerical precision (see Table~\ref{tab:summary}).
Consequently, for each impurity orbital and spin, the differences in the integrated removal and addition weights obey
\begin{equation}
  \Delta\!\sum_n \wminus{n}{a\sigma}
  +
  \Delta\!\sum_m \wplus{m}{a\sigma}
  =
  0,
  \label{eq:weight_difference}
\end{equation}
where $\Delta$ denotes the QSCI value minus the ED value.
The removal-weight sum equals the corresponding ground-state occupation, so the integrated removal-weight difference is equal to the occupation difference discussed in Sec.~\ref{sec:results:occ}.
Thus, deviations in the integrated weights of the two channels are equal and opposite, while the frequency-resolved differences need not be.

\subsection{Self-energy}
\label{sec:results:sigma}

The impurity self-energy $\Sigma_{\mathrm{imp}}(z)$ is a $5\times5$ matrix, which we obtain from Eq.~\eqref{eq:SigmaImp} for ED and from Eq.~\eqref{eq:SigmaFull} below for QSCI.
For the present model, the off-diagonal self-energy components are small compared with the diagonal components, with a relative magnitude of order $1\%$ over the range examined.
On the Matsubara axis, $\operatorname{Re}\Sigma_{aa}(\ii\omega_n)$ approaches its high-frequency Hartree limit.
At the largest Matsubara frequency considered, the difference is within $0.02\%$ for every orbital.

\begin{figure*}[t]
\centering
\includegraphics[width=\textwidth]{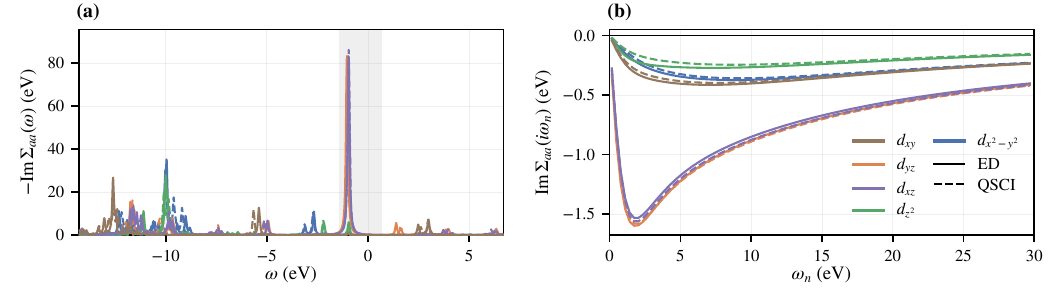}
\caption{\label{fig:sigma}%
Imaginary parts of the diagonal impurity self-energy components for the five impurity orbitals, obtained using exact diagonalization (solid lines) and QSCI (dashed lines).
(a) $-\operatorname{Im}\Sigma_{aa}(\omega+\mathrm{i}\eta)$
on the real-frequency axis, with frequencies measured
relative to the chemical potential and broadening
$\eta=0.05$~eV.
(b) $\operatorname{Im}\Sigma_{aa}(\mathrm{i}\omega_n)$
at Matsubara frequencies for
$\beta=20$~eV$^{-1}$.
}
\end{figure*}

The Weiss Green's function is constructed from the same finite-bath parameters used in the AIM Hamiltonian:
\begin{align}
  \bigl[\mathcal{G}_{0,\mathrm{AIM}}^{-1}(z)\bigr]_{ab}
  &=
  (z+\chempot)\delta_{ab}
  -\epsimp{ab}
  -[\Delta_{\mathrm{AIM}}(z)]_{ab},
  \nonumber\\
  [\Delta_{\mathrm{AIM}}(z)]_{ab}
  &=
  \sum_{\ell}
  \frac{
    V_{a\ell}V_{b\ell}^{*}
  }{
    z+\chempot-\epsbath{\ell}
  }.
  \label{eq:weiss}
\end{align}
Here $z=\omega+\ii\broad$ for the broadened real-frequency calculation and $z=\ii\omega_n$ on the Matsubara axis.
Using the same chemical potential and energy convention for $\mathcal{G}_{0,\mathrm{AIM}}$ and $G_{\mathrm{imp}}$ ensures consistency in the Dyson equation.

Figure~\ref{fig:sigma} compares the diagonal imaginary parts of the QSCI and ED self-energies on the real-frequency and Matsubara axes.
Table~\ref{tab:sigma} lists the ED values of $\operatorname{Re}\Sigma_{aa}(0)$ and $\operatorname{Re}\Delta_{aa}(0)$, together with the corresponding diagonal impurity one-body terms.
The impurity levels near $-26$~eV are largely compensated by $\operatorname{Re}\Sigma_{aa}(0)$, and the sums are lowest for $d_{x^{2}-y^{2}}$ and $d_{z^{2}}$, the two nearly doubly occupied orbitals.

\begin{table}[t]
\caption{\label{tab:sigma}%
Diagonal impurity one-body terms and real parts of the self-energy and hybridization function, in eV, for the finite-bath AIM using exact diagonalization.
The self-energy and hybridization function are evaluated at $z=0+\ii\broad$, with $\broad=0.3$~eV; we write $\Sigma_{aa}(0)$ and $\Delta_{aa}(0)$ for these values.
The final row gives their sum with the impurity one-body term, $\varepsilon^{\mathrm{imp}}_{aa} +\operatorname{Re}\Sigma_{aa}(0) +\operatorname{Re}\Delta_{aa}(0)$.
The impurity one-body terms include the double-counting correction, as in Table~\ref{tab:aimparams}.
}
\begin{ruledtabular}
\setlength{\tabcolsep}{3pt}
\begin{tabular}{lccccc}
 & $d_{x^{2}-y^{2}}$ & $d_{yz}$ & $d_{z^{2}}$ & $d_{xz}$ & $d_{xy}$ \\
\colrule
$\epsimp{aa}$         & $-25.810$ & $-25.445$ & $-25.820$ & $-25.466$ & $-26.384$ \\
$\Real\Sigma_{aa}(0)$ & $23.023$  & $25.683$  & $22.868$  & $26.096$  & $23.456$ \\
$\Real\Delta_{aa}(0)$ & $0.074$   & $-0.265$   & $0.379$   & $0.071$   & $7.130$ \\
\colrule
Sum                   & $-2.713$  & $-0.027$  & $-2.573$  & $+0.701$  & $+4.202$ \\
\end{tabular}
\end{ruledtabular}
\end{table}

For the exact Green's function, the poles of $\Delta_{\mathrm{AIM}}$ in Eq.~\eqref{eq:SigmaImp} are canceled by poles of $G_{\mathrm{imp}}^{-1}$.
For a Green's function evaluated within selected subspaces, this cancellation is not exact in general, and Eq.~\eqref{eq:SigmaImp} does not guarantee a causal self-energy.
Related difficulties have been reported for truncated-space impurity
solvers~\cite{Go2017}.

We therefore obtain the QSCI self-energy by applying the Dyson equation to the impurity and bath orbitals together~\cite{Zhu2019}:
\begin{equation}
  \Sigma_{\mathrm{full}}(z)
  =
  (z+\chempot)I-\varepsilon^{\mathrm{imp}}-\Delta_{\mathrm{AIM}}(z)-\widetilde{G}_{\mathrm{imp}}^{-1}(z).
  \label{eq:SigmaFull}
\end{equation}%
Here $\widetilde{G}_{\mathrm{imp}}^{-1}(z)=[G_{\mathrm{full}}^{-1}(z)]_{\mathrm{imp}}-\Delta_{\mathrm{AIM}}(z)$, $G_{\mathrm{full}}$ is the $10\times10$ Green's function of the operators $\hat{a}_{p\sigma}$, defined as in Eq.~\eqref{eq:lehmann}, and $[\cdot]_{\mathrm{imp}}$ denotes the $5\times5$ impurity block.
Because the bath orbitals are noninteracting, $\widetilde{G}_{\mathrm{imp}}=G_{\mathrm{imp}}$ for the exact Green's function, and Eq.~\eqref{eq:SigmaFull} then coincides with Eqs.~\eqref{eq:SigmaImp} and~\eqref{eq:weiss}.
The augmentation in Eq.~\eqref{eq:aug} includes all ten orbitals, so the QSCI $G_{\mathrm{full}}$ also satisfies the sum rule; all poles of $\Sigma_{\mathrm{full}}$ then have positive-semidefinite residues, and $\operatorname{Im}[\Sigma_{\mathrm{full}}(z)]_{aa}\leq0$ throughout the upper half plane.
For QSCI, $\widetilde{G}_{\mathrm{imp}}$ differs from $G_{\mathrm{imp}}$; their total spectral functions differ by $0.019$~eV$^{-1}$, measured as in Eq.~\eqref{eq:spectral_error}, compared with $\epsilon_A=0.035$~eV$^{-1}$.
In Figs.~\ref{fig:sigma} and~\ref{fig:gap}(b), the QSCI self-energy is $\Sigma_{\mathrm{full}}$.

\subsection{Charge-transfer gap}
\label{sec:results:gap}

\begin{figure*}[t]
\centering
\includegraphics[width=\textwidth]{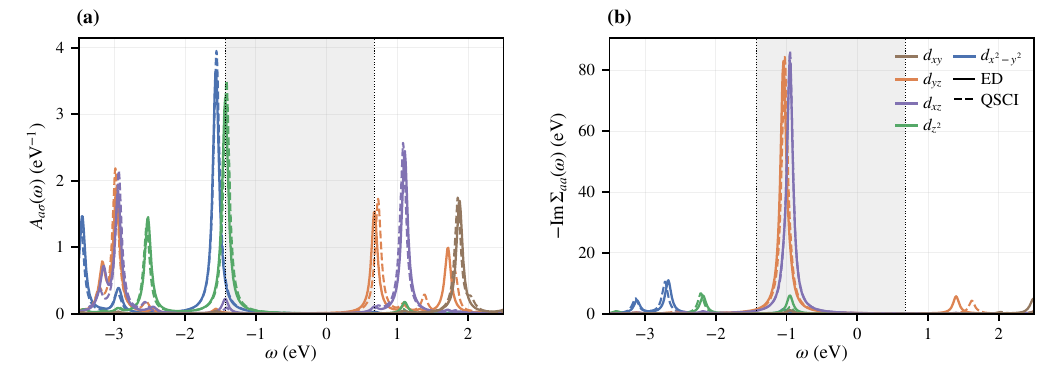}
\caption{\label{fig:gap}%
Impurity spectral function and self-energy near the charge gap, obtained using exact diagonalization (solid lines) and QSCI (dashed lines).
Frequencies are measured relative to the chemical potential.
(a) Spin-resolved orbital spectral functions $A_{a\sigma}(\omega)$.
(b) $-\operatorname{Im}\Sigma_{aa}(\omega+\mathrm{i}\eta)$.
Both panels use $\eta=0.05$~eV.
Dotted lines mark the ED particle-removal and particle-addition thresholds, $\omega=-1.431$ and $+0.677$~eV, and the shaded region is the charge gap.
}
\end{figure*}

We quantify the gap in the impurity spectrum [Fig.~\ref{fig:spectral}(b)] by the charge gap of the finite AIM,
\begin{equation}
  \Delta_{c}
  =
  E_{0}^{N+1}+E_{0}^{N-1}-2E_{0}^{N},
  \label{eq:gap}
\end{equation}%
with $E_{0}^{N\pm1}$ as in Eq.~\eqref{eq:muwindow}.
ED gives $\Delta_{c}=2.108$~eV, with the particle-removal and particle-addition thresholds at $\omega=-1.431$ and $+0.677$~eV, respectively.
QSCI gives $\Delta_{c}=2.142$~eV, $34$~meV larger, with thresholds at $-1.413$ and $+0.729$~eV.

Both thresholds have mainly impurity character.
Within $0.5$~eV of the removal and addition thresholds, the impurity orbitals carry $93\%$ and $82\%$, respectively, of the Lehmann weight of $\hat{a}_{p\sigma}$ and $\hat{a}^{\dagger}_{p\sigma}$ summed over all orbitals $p$, mainly in $d_{x^{2}-y^{2}}$ and $d_{z^{2}}$ for removal and in $d_{yz}$ and $d_{xz}$ for addition [Fig.~\ref{fig:gap}(a)].
QSCI gives $93\%$ and $79\%$, respectively.

Figure~\ref{fig:gap}(b) shows the corresponding self-energies.
Inside the gap, the ED self-energy has two poles, at $\omega=-1.044$ and $-0.952$~eV, with residues of $4.1$~eV$^{2}$ in the $d_{yz}$ and $d_{xz}$ channels, respectively, and at most $0.3$~eV$^{2}$ in the other orbitals.
QSCI reproduces both poles, at $-1.026$ and $-0.951$~eV, with residues within $5\%$ of the ED values.

Coming back to the heme molecule within DFT+DMFT, the total DOS shown in Fig.~\ref{fig:spectral}(a) exhibits a charge-transfer gap between the N $2p$ and Fe $3d$ bands.
The highest occupied and lowest unoccupied peaks are situated at $-0.72$ and $+0.76$~eV, respectively, $1.48$~eV apart.
The N $2p$ states are placed between the lower and the upper Hubbard bands of the Fe $3d$ states, presenting a typical charge-transfer insulator picture~\cite{ZSA1985}.
To the best of our knowledge, this is the first time that the gap of heme has been identified as a charge-transfer gap and computed beyond DFT; the DFT estimate of the gap, $0.32$~eV~\cite{Cha2025}, is significantly smaller. 

\subsection{Convergence of the spectral function}
\label{sec:results:conv}

The spectral error converges as the number $M$ of retained eigenstates increases.
Retaining only a few hundred eigenstates or fewer, however, introduces a spectral error comparable to the deviation from ED obtained using all eigenstates.
The spectral error decreases monotonically as the determinant cap is raised, and we therefore use the uncapped subspaces for the results reported here.
Appendix~\ref{app:conv} presents these convergence tests and examines the dependence on the broadening parameter $\broad$.

\section{Discussion}
\label{sec:discussion}

\subsection{Role of the quantum processor}
\label{sec:disc:role}

In the present workflow, the quantum processor prepares a trial state and samples its computational-basis measurement outcomes.
The retained bit strings identify Slater determinants that define the ground-state subspace after postselection and augmentation under time reversal.
The Hamiltonian matrix elements and ground-state expansion coefficients are then computed classically.
The particle-addition and particle-removal subspaces, Green's function, and self-energy are also constructed through classical post-processing.

Finite sampling and device noise affect the selected determinants and therefore the accuracy of the resulting ground state and spectral quantities.
Nevertheless, for any nonempty selected subspace within the target sector, classical diagonalization of the projected Hamiltonian yields a variational upper bound to the ground-state energy in that sector.
This separation between quantum sampling and classical diagonalization allows the calculation to use noisy measurement outcomes without estimating the energy directly on the quantum processor.
The bound alone, however, does not guarantee that the selected subspace is sufficiently accurate.

The performance of the method depends on the prepared state and the determinants sampled from it.
As discussed in Sec.~\ref{sec:results:gs}, some configurations with significant weight in the exact ground state are absent from the selected subspace.
Methods that enrich the sampled subspace by evolving or adaptively constructing the input state~\cite{Mikkelsen2025,Sugisaki2024,yu2025,Nakagawa2024} may improve the representation of such configurations.
In the present demonstration, the trial state is a single Hartree--Fock determinant; the results therefore establish the accuracy of the hybrid workflow for this impurity model, without demonstrating a quantum computational advantage.

\subsection{Toward fault-tolerant impurity solvers}
\label{sec:disc:ft}

The Lehmann representation in Sec.~\ref{sec:method:lehmann} is independent of the method used to obtain the ground-state energy and the particle-addition and particle-removal spectral information.
In the present work, these quantities are obtained through classical diagonalization in the subspaces constructed from quantum measurements.
Fault-tolerant quantum algorithms may provide alternative routes to the same quantities.
For example, quantum phase estimation (QPE)~\cite{Kitaev1995,Abrams1999,AspuruGuzik2005} can be used to estimate excitation energies and, with suitable state preparation and measurements, spectral weights.
Quantum singular value transformation (QSVT)~\cite{Gilyen2019,Low2019} also provides a framework for constructing algorithms to evaluate Green's functions.
These approaches need not explicitly enumerate all eigenstates in the $N\pm1$ sectors.

The interface with classical post-processing depends on the quantum algorithm.
In QSCI, the eigenvectors are represented classically in the selected determinant bases.
The transition matrix elements are then evaluated from the action of the fermionic operators on those determinants and overlaps with the $N\pm1$ eigenstates.
In a fault-tolerant approach, relevant states and operators may instead be represented on quantum registers, requiring additional quantum operations and measurements to extract spectral information.
Once the required Green's-function matrix elements have been obtained, the self-energy can be evaluated through the same Dyson equation used here.
Comparing the costs of these approaches requires consideration of state preparation, spectral resolution, measurement precision, and classical post-processing, as well as the size of the impurity model.

State preparation remains important in both settings, although its role differs.
In QPE, the probability of obtaining a target eigenstate depends on the overlap of the prepared state with that eigenstate.
In QSCI, the prepared state's computational-basis distribution determines the probability of sampling each determinant.
Improving these overlaps or sampling probabilities may therefore be important for extending quantum impurity solvers to larger and more strongly correlated systems.

\section{Conclusion}
\label{sec:conclusion}

We have applied quantum-selected configuration interaction (QSCI) to an Anderson impurity model constructed for the heme $b$ cofactor of hemoglobin.
The model comprises the five Fe $3d$ orbitals and five bath orbitals and is encoded on $20$ qubits of the Quantinuum Reimei trapped-ion processor.
The selected-subspace diagonalization and subsequent Green's-function and self-energy calculations are performed classically.

Quantum sampling, followed by postselection and augmentation under time reversal, yields a ground-state subspace of $444$ determinants, spanning $1.0\%$ of the target particle-number sector.
Diagonalization within this subspace gives a ground-state fidelity of $0.982$ and an energy approximately $49$~meV above the exact
diagonalization (ED) reference.
Using classically constructed particle-addition and particle-removal subspaces, we obtain the real-frequency spectral function from the Lehmann representation without analytic continuation.
The resulting spectrum is in good quantitative agreement with ED, and the corresponding Green's function yields the impurity self-energy through the Dyson equation for the impurity and bath orbitals together, which makes it causal by construction. QSCI also recovers the spin-triplet ground state, reproduces the approximately $2.1$~eV impurity spectral gap to within $34$~meV, and captures both in-gap self-energy poles.
Finally, based on our calculations, we identify heme as a charge-transfer insulator with a gap of $1.48$~eV. 

The Lehmann representation is independent of the method used to obtain the relevant spectral information, providing a connection to future fault-tolerant approaches.
A natural next step is to incorporate a quantum impurity solver into the DMFT self-consistency cycle and examine how errors in the Green's function and self-energy affect its convergence and the resulting physical observables.

\begin{acknowledgments}
This work is partially based on results obtained from a project commissioned by the New Energy and Industrial Technology Development Organization (NEDO) (Project No.~JPNP20017).
K.~S. acknowledges support from JSPS KAKENHI Grant No.~JP26K06954.
S.~Y. acknowledges support from JSPS KAKENHI Grant No.~JP24K02948.
S.~Y. also acknowledges support from JST COI-NEXT Grant No. JPMJPF2221, the UTokyo Quantum Initiative, the RIKEN TRIP initiative (RIKEN Quantum), and the COE Research Grant in Computational Science from Hyogo Prefecture and Kobe City through the Foundation for Computational Science.
This work used computational resources of the supercomputer Fugaku provided by the RIKEN Center for Computational Science (Project ID: ra010014).
\end{acknowledgments}

\appendix

\section{Impurity-model parameters and convergence tests}
\label{app:conv}

\subsection*{Impurity-model parameters}

The Anderson impurity model is specified by the one-body Hamiltonian in the space of ten spatial orbitals and the interaction tensor within the five impurity orbitals.
The diagonal one-body parameters are listed in Table~\ref{tab:aimparams}; additional information about the model and the spectral calculations is given below.

The impurity--bath hybridization amplitudes $V_{a\ell}$ form a real $5\times5$ matrix, with the largest magnitude of order $4$~eV; $11$ of the $25$ elements exceed $0.1$~eV in magnitude.
The three largest are $V_{d_{xy},2} = 3.646$~eV, $V_{d_{xy},0} = 2.389$~eV, and $V_{d_{z^{2}},1} = 1.662$~eV, coupling to bath orbitals at $-1.915$, $-16.171$, and $-6.402$~eV, respectively.

The interaction tensor includes pair-dependent exchange integrals $J_{ab}$.
We examined the effect of approximating the full tensor.
Restricting it to the Kanamori form, which retains only the density-density, spin-flip, and pair-hopping terms, raises the ground-state energy by $0.397$~eV ($0.25\%$).
Additionally replacing the pair-dependent exchange integrals by their average gives $0.422$~eV ($0.27\%$), and a density-density interaction gives $0.722$~eV ($0.45\%$).
These shifts are several times larger than the QSCI energy error.
The interaction tensor was also verified to commute with the total spin operator $\hat{S}^{2}$.

Unless otherwise stated, spectral functions are evaluated on a uniform grid of $1001$ points over $-20\leq\omega\leq6$~eV, using a broadening of $\broad=0.3$~eV and chemical potential $\chempot=0$.
The spectral error of Eq.~\eqref{eq:spectral_error} is evaluated on this grid and over this interval throughout.

\subsection*{Convergence of the classical post-processing}

The classical evaluation of the spectral function allows two numerical truncations: a limit $\Mkeep$ on the number of eigenstates retained in each adjacent sector in Eq.~\eqref{eq:lehmann}, and a cap on the number of determinants retained in the particle-removal subspaces $\SNm$.
Both settings can be varied using the same measurement data, without additional executions on the quantum processor.

\begin{figure*}[t]
\centering
\includegraphics[width=\textwidth]{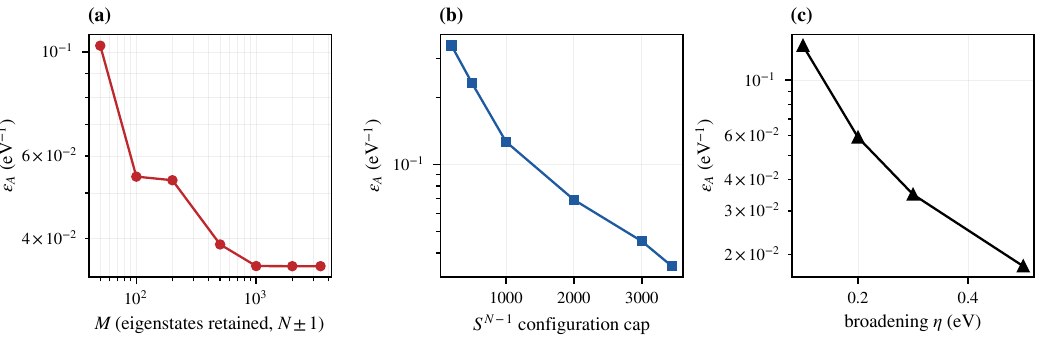}
\caption{\label{fig:conv}%
Dependence of the spectral error on (a) the maximum number of eigenstates $\Mkeep$ retained in each adjacent sector,
(b) the cap on the number of determinants in the particle-removal subspaces $\SNm$, and
(c) the broadening $\broad$.
The main-text results retain all eigenstates within the augmented particle-addition and particle-removal subspaces, impose no additional determinant cap, and use $\broad=0.3$~eV.
}
\end{figure*}

Figure~\ref{fig:conv}(a) shows the dependence of
the spectral error on $\Mkeep$. As $\Mkeep$ increases, the error decreases over the range examined and becomes nearly converged
beyond $\Mkeep\simeq1000$.
The change between this value and the result obtained using all eigenstates within the augmented subspaces is $0.05\%$.
At $\Mkeep=500$, the error is $11\%$ larger.
Restricting the number of retained eigenstates further increases the spectral error relative to ED.

Figure~\ref{fig:conv}(b) shows the dependence on the determinant cap.
The error decreases monotonically as the cap is raised, from $0.34$~eV$^{-1}$ at a cap of $200$ to $0.045$~eV$^{-1}$ at $3000$, and reaches $0.035$~eV$^{-1}$ without a cap.
Truncation errors can be comparable to, or larger than, the difference between the uncapped QSCI and ED spectra.
It is therefore important to specify the classical truncation settings when assessing the accuracy of a sampling-based impurity solver.

Figure~\ref{fig:conv}(c) shows the dependence of the spectral error on the broadening parameter.
The error decreases from $1.4\times10^{-1}$ at $\broad=0.1$~eV to $3.5\times10^{-2}$ at $\broad=0.3$~eV, the value used in the main text, and to $1.8\times10^{-2}$ at $\broad=0.5$~eV.
Increasing $\broad$ smooths differences between the discrete spectral features of the QSCI and ED calculations.
The ratio of the particle-removal and particle-addition errors remains approximately stable over this range, between $1.4$ and $1.6$.
Thus, the difference between the two channel errors discussed in Sec.~\ref{sec:results:spectral} persists for the broadening values examined.

\bibliography{references}

\end{document}